\documentstyle [12pt]{article}
\advance\voffset by -1 in
\advance\hoffset by -.5 in
\def\d{\delta}

\def\l{\lambda}

\def\p{\partial}

\def\O{\Omega}
\def\Om0{\Omega^0}
\def\O1{\Omega^1}
\def\cA{{\cal A}}

\def\tf{\tilde{f}}
\def\tg{\tilde{g}}

\def\res{{\rm res}\:}

\def\cL{{\cal L}}
\def\cR{{\cal R}}

\def\al{(A^{-1}[P_l,A])_-}
\def\ak{(A^{-1}[P_k,A])_-}
\def\Lra{\Leftrightarrow}
\begin{document}

\centerline{{\Large\bf On the constrained KP hierarchy II}}

\vspace{.4in}
\centerline{{\bf L.A.Dickey}}

\vspace{.2in}
\centerline{Dept. Math., University of Oklahoma, Norman, OK 73019\footnote{
Internet: "ldickey@nsfuvax.math.uoknor.edu"}}

\bigskip
\centerline{{\bf Abstract}}

\vspace{.1in}
\small
A constrained KP hierarchy is discussed that was recently suggested by
Aratyn et al. and by Bonora et al. This hierarchy is a restriction of the
KP to a submanifold of operators which can be represented as a ratio of two
purely differential operators of prescribed orders. Explicit formulas for
action of vector fields on these two differential operators are written which
gives a new description of the hierarchy and provides a new, more constructive,
proof of compatibility of the constraint with the hierarchy. Also the Poisson
structure of the constrained hierarchy is discussed.
\normalsize

\vspace{.3in}
In the second part of this paper (see [3]), a new type of a constrained KP
hierarchy is discussed that was recently suggested by Aratyn et al. [1] and
Bonora et al. [2]. We will use the version of this hierarchy given in [2].

As it is well-known, an equivalent description of the KP hierarchy can be
obtained in terms of an pseudo-differential operator of the $n$th order,
instead of the first one. A new constrained hierarchy is a restriction of the
KP to the submanifold of operators which can be
represented as ratios of two purely differential operators of given orders,
say, $A$ of the $(n+m)$th and $B$ of the $m$th order. Thus, $L=AB^{-1}$.
This restricted hierarchy was called the $(n,m)$th hierarchy in [2].

First of all, it must be proven that this submanifold is compatible with the KP
hierarchy, i.e., vector fields representing the equations of the hierarchy are
tangent to the submanifold. This was done in [1] and [2]. Operators belonging
to this restriction
depend on finite number of fields, namely, coefficients of the operators $A$
and $B$. One can find how the vector fields of the hierarchy act on these
generators, i.e., how they act on operators $A$ and $B$ separately. Thus, an
alternative description of the hierarchy arises in terms of pairs of
differential operators $A,B$ and differential equations having a form $$\p_{t_k
}A=f(A,B),~~\p_{t_k}B=g(A,B).\eqno{(1)}$$

{\em The goal of this note is to point out explicit formulas for $f$ and $g$
in Eq. (1) (see below Eq.(3)), then to prove that all vector fields commute
(Proposition 2 below), and that thus constructed
hierarchy can be embedded into KP by the relation $L=AB^{-1}$ (Proposition 1
below). This gives another, constructive, proof of the compatibility
of the constraint with the KP hierarchy.} The Poisson structure of the
constrained hierarchy also is discussed.

Notice that the operator $AB^{-1}$ can be written in a different form using
the fact that for every differential operator $K$ and a first-order operator
$\p-S$ there exists a division with a remainder term: $K=M(\p-S)+f$ where $M$
is another differential operator and $f$ is a function. Then, factorizing $B$:
$B=(\p-S_1)...(\p-S_m)$, it is easy to find that $AB^{-1}$
can be represented as $$L=\p^n+\sum_{l=0}^{n-1}a_l\p^{n-l-1}+\sum_{l=1}^ma_{n+l
-1}(\p-S_l)^{-1}...(\p-S_1)^{-1}\eqno{(2)}$$ (see [2], the hierarchy in a
similar form also was introduced in [1]). However, we do not use this
representation in what follows.

We have the following notations. Let $t_k$ be some variables, $\p_k$
be derivations with respect to these variables, $A$ and $B$ differential
operators,
$$A=\p^{n+m}+a_1\p^{n+m-1}+a_2\p^{n+m-2}+...+a_{n+m},~~B=\p^m+b_1\p^{m-1}+
b_2\p^{m-2}+...+b_m.$$ Let $$P_k=((AB^{-1})^{k/n})_+$$ where $k$ is a positive
integer. Subscripts $+$ and $-$ mean, as usual, the non-negative and the
negative parts of a pseudo-differential operator.

{\em Define the equations of the hierarchy as} $$\p_kA=A(A^{-1}[P_k,A])_-,~~
\p_kB=B(B^{-1}[P_k,B])_-.\eqno{(3)}$$ The equations also can be written as
$$\p_kA=[P_k,A]-A(A^{-1}[P_k,A])_+,~\p_kB=[P_k,B]-B(B^{-1}[P_k,B])_+.\eqno{(3a)}
$$ The equations are well defined since the Eqs. (3) and (3a) show that their
right-hand sides are purely differential operators of orders less than those of
$A$ and, correspondingly, $B$.\\

{\bf Remark.} The mapping $(A,B)\mapsto AB^{-1}$ yields an embedding of the
differential algebra $\cA_{L}$ which is generated by coefficients of the
operator $L$ into the differential algebra $\cA_{A,B}$ comprising differential
polynomials in coefficients $\{a_i\}$ and $\{b_i\}$. The algebra $\cA_{A,B}$
is considerably larger then $\cA_L$. \\

{\bf Proposition 1.} {\sl The equations (3) imply
$$\p_k(AB^{-1})=[P_k,AB^{-1}],
\eqno{(4)}$$i.e., the operator $L=AB^{-1}$ satisfies the equations of the KP
hierarchy.}\\

{\em Proof.} First, we notice that the order of the operator $[P_k,AB^{-1}]$ is
less than that of the operator $AB^{-1}$. Indeed, $${\rm ord}~[P_k,AB^{-1}]=
{\rm ord}~[(AB^{-1})_+^{k/n},AB^{-1}]={\rm ord}~[AB^{-1},(AB^{-1})_-^{k/n}]<
{\rm ord}~AB^{-1}-1.$$ We have
$$[P_k,AB^{-1}]=[P_k,A]B^{-1}-AB^{-1}[P_k,B]B^{-1}$$ $$=A(A^{-1}[P_k,A]-B^{-1}
[P_k,B])B^{-1}.$$The operator in the parentheses is of a negative order,
otherwise the order of $[P_k,AB^{-1}]$ would be not less than that of
$AB^{-1}$. Hence nothing will change if we write this operator with a subscript
$-$. Now, $$[P_k,AB^{-1}]=A(A^{-1}[P_k,A]-B^{-1}
[P_k,B])_-B^{-1}$$ $$=A((A^{-1}[P_k,A])_--(B^{-1}[P_k,B])_-)B^{-1}.$$ On the
other hand, $$\p_k(AB^{-1})=(\p_kA)B^{-1}-AB^{-1}(\p_kB)B^{-1}$$ $$=
A(A^{-1}[P_k,A])_-B^{-1}-AB^{-1}(B(B^{-1}[P_k,B])_-)B^{-1}$$ $$=
A((A^{-1}[P_k,A])_--(B^{-1}[P_k,B])_-)B^{-1},$$ i.e., the same expression.
$\Box$ The equality (4) yields, as usual:\\

{\bf Corollary.} {\sl The equations $$\p_lP_k-\p_kP_l=[P_l,P_k]\eqno{(5)}$$
hold.}\\

{\bf Proposition 2.} {\sl Vector fields $\p_k$ commute.}\\

{\bf Remark.} They commute in their action on $AB^{-1}$ as restrictions of
commuting vector fields. This, however, does not necessarily mean that they
commute on the generators, i.e., on $A$ and $B$.

{\em Proof.}$$\p_l\p_kA-\p_k\p_lA=\p_l(A\ak) -\p_k(A\al)=(\p_lA)\ak$$ $$-
A(A^{-1}(\p_lA)A^{-1}[P_k,A])_-+A(A^{-1}[\p_l P_k,A])_-+A(A^{-1}[P_k,\p_lA])_-
-(k\Lra l)$$ $$=A\{\al\cdot\ak-\al A^{-1}[P_k,A]$$ $$+A^{-1}[\p_l P_k,A]+
A^{-1}[P_k,A\al]\}_--(k\Lra l).$$ We are going to prove
that this is zero, for this reason the common factor $A$ is irrelevant and can
be omitted. Let $a^k=A^{-1}[P_k,A]$ and $a^l=A^{-1}[P_l,A]$.
We have a sum of four terms: $$ (I)=[a_-^l,a_-^k]=[a_-^l,a_-^k]_-;~
(II)=-(a_-^la^k)_-+(a_-^ka^l)_-;$$ $$(III)=(A^{-1}[P_k,Aa_-^l])_--(A^{-1}[P_l,A
a_-^k])_-;$$ $$(IV)=(A^{-1}[\p_lP_k-\p_kP_l,A])_-=(A^{-1}[[P_l,P_k],A])_-.$$ In
the last transformation we have used (5). Now,$$(I)+(II)=(-a_-^la_+^k+a_-^ka_+^
l)_-=(-a^la_+^k+a^ka_+^l)_-.$$ We transform $(III)$:
$$(III)=(A^{-1}[P_k,A]a_-^l)_--(A^{-1}[P_l,A]a_-^k)_-+[P_k,a_-^l]_--[P_l,a_-^k]
_- $$ $$=(a^ka_-^l-a^la_-^k)_-+([P_k,a^l]-[P_l,a^k])_-=(IIIa)+(IIIb).$$ Then,
$$(I)+(II)+(IIIa)=[a^k,a^l]_-;$$ $$(IIIb)=[P_k,A^{-1}[P_l,A]]_--(k\Lra l)=
([P_k,A^{-1}][P_l,A]+A^{-1}[P_k,[P_l,A]])_--(k\Lra l)$$ $$=(-A^{-1}[P_k,A]A^{-1
}[P_l,A]-(k\Lra l))_-+(A^{-1}[[P_k,P_l],A])_-$$ $$=-[a^k
,a^l]_-+(A^{-1}[[P_k,P_l],A])_-.$$ The first term cancels with
$(I)+(II)+(IIIa)$
and the second with $(IV)$ that completes the proof. Replacing $A$ by $B$ in
this proof, we obtain that the vector fields also commute on $B$. $\Box$

The proven fact entitles us to call the equations (3) a hierarchy of integrable
equations since all the equations have infinitely many symmetries, every
equation provides a symmetry for all others.\\

We also discuss a problem of Poisson structure of the constrained hierarchy.
There are infinitely many pairs of compatible Poisson structures for the
general KP hierarchy (see, e.g., [4]). In every pair a certain, always the
same, structure has a nickname ``the first", the other is ``the second". If the
KP hierarchy is expressed in terms of a pseudo-differential operator $L=\p^n
+u_1\p^{n-1}+u_2\p^{n-2}+...$ of an order $n$, then one of the pairs is in a
sense natural. That is one which transforms to
the $n$th KdV structure under the restriction $L=L_+$. As we will see, the
``second" structure of this pair can be restricted to our constrained
hierarchy. Let $\tf=\int fdx$ and $\tg=\int gdx$ be two functionals
where $f$ and $g$ are differential polynomials in coefficients of the operator
$L$. Then the second Poisson bracket is $$\{\tf,\tg\}=\int\res [({\d f\over
\d L}L)_-{\d g\over\d L}L-(L{\d f\over\d L})_-L{\d g\over\d L}]dx \eqno{(6)}$$
where $\d f/\d L=\sum_{i=1}^\infty\p^{-n+i-1}\d f/\d u_i$ and, similarly, for $
\d g/\d L$. (The first structure can be obtained from the second one as a
coefficient in $\l$ if $L$ is replaced by $L+\l$ in the above formula).

Let $\cL$ be the manifold of all pseudo-differential $n$th order operators, and
$\cR$ the submanifold of operators of the form $L=AB^{-1}$ with $(n+m)$th order
differential operator $A$ and $m$th order $B$. The functionals $\tf$ and $\tg$
are given only on $\cR$ now. In order to restrict the Poisson bracket to $\cR$
one must prolong the functionals from the submanifold to the whole manifold and
take the Poisson bracket there. If the result is independent of the
continuation, it determines the restriction of the Poisson bracket to the
submanifold. In what follows, it will be proven that this is exactly the case.

Let $A=\sum_{i=0}^{n+m}a_i\p^{n+m-i}$ and $B=\sum_{i=0}^mb_i\p^{m-i}$ where
$a_0=b_0=1$. Let $${\d f\over\d A}=\sum_{i=1}^{n+m}\p^{-n-m+i-1}{\d f\over\d a_
i}~{\rm and }~{\d g\over\d B}=\sum_{i=1}^{m}\p^{-m+i-1}{\d g\over\d b_i}.$$
Now, let us take a variation of $\tf$ along the submanifold: $$\d\tf=\int\res
{\d f\over\d L}\d Ldx=\int\res{\d f\over\d L}\d(AB^{-1})dx$$ $$=\int\res
{\d f\over\d L}(\d A\cdot B^{-1}-AB^{-1}\d B\cdot B^{-1})dx=\int\res(B^{-1}
{\d f\over\d L}\cdot\d A-B^{-1}{\d f\over\d L}AB^{-1}\cdot\d B)dx$$ whence
$${\d f\over\d A}=(B^{-1}{\d f\over\d L})_-,~{\d f\over\d B}=-(B^{-1}
{\d f\over\d L}AB^{-1})_-.$$ Further, Eq.(6) implies $$\{\tf,\tg\}=
\int\res[AB^{-1}({\d f\over\d L}AB^{-1})_--(AB^{-1}{\d f\over\d L})_-AB^{-1}]
{\d g\over\d L}dx.$$ Computation of variational derivatives $\d f/\d L$ and
$\d g/\d L$ requires a continuation of functionals $\tf$ and $\tg$ from $\cR$
to the whole $\cL$. However, the above expression can be rewritten as
$$\{\tf,\tg\}=-\int\res[AB^{-1}(B{\d f\over\d B})_-+(A{\d f\over\d A}
)_-AB^{-1}]{\d g\over\d L}dx,\eqno{(7)}$$ i.e., $\d f/\d L$ is eliminated and
expressed in terms of $\d f/\d A$ and $\d f/\d B$ which no more depend on the
continuation of the functional $\tf$ since $A$ and $B$ are inner variables.
The bracket is antisymmetric with respect to $f$ and $g$, thus, it does not
depend on the continuation of the functional $\tg$ either.

Unfortunately, we cannot express this bracket only in terms of $\d f/\d A,
{}~\d f/\d B, ~\d g/\d A$ and $\d g/\d B$. Perhaps, this means that it is not
local in terms of generators $A$ and $B$.

Let us write the equation (4) in a Hamiltonian form with respect to the above
structure. We need the following notations. Let $R$ be the algebra of all
pseudo-differential operators, $R_+$ the subalgebra of differential operators,
and $R_-$ that of integral operators. We have $L\equiv AB^{-1}\in\p^{n+1}R_-$.
The variational derivative $\d f/\d L$ can be understood as element of the
quotient $R/\p^{-n}R_-$, so $\d f/\d L=\sum_{i=1}^\infty\p^{-n+i-1}\d f/\d u_i$
mod $\p^{-n}R_-$. It can be proven, see [4], Proposition 6.3.2, that if
$\tf=(n/k)\int\res L^{k/n}dx$ then $\d f/d L=L^{(k-n)/n}$ mod $\p^{-n}R_-$.
Now, the right-hand side of (4) is $$[L_+^{k/n},L]=(LL^{(k-n)/n})_+L-
L(L^{(k-n)/n}L)_+$$ $$=(L(L^{(k-n)/n}|_{{\rm mod}~\p^{-n}R_-}))_+L-
L(L^{(k-n)/n}|_{{\rm mod}~\p^{-n}R_-})L)_+$$ $$=(L{\d f\over\d L})_+L-L(
{\d f\over\d L}L)_+=(AB^{-1}{\d f\over\d L})_+AB^{-1}-AB^{-1}({\d f\over\d L}
AB^{-1})_+$$ $$=-(AB^{-1}{\d f\over\d L})_-AB^{-1}+AB^{-1}({\d f\over\d L}
AB^{-1})_-=-(A{\d f\over\d A})_-AB^{-1}-AB^{-1}(B{\d f\over\d B})_-.$$ The
equation (4) takes the form $$\p_k(AB^{-1})=-(A{\d f\over\d
A})_-AB^{-1}-AB^{-1}
(B{\d f\over\d B})_-.\eqno{(8)}$$ This is a Hamiltonian equation with respect
to the brackets (7) with the hamiltonian $\tf=(n/k)\int\res L^{k/n}dx$. The
variational derivatives are evaluated only with respect to the inner to the
submanifold $\cR$ variables $A$ and $B$.\\

{\bf Discussion.} We believe that, irrespective of embedding into KP, the
integrable hierarchy (3) defined on pairs of differential operators is
interesting in its own right. Does it have Poisson, or Hamiltonian, or, still
better, bi-Hamiltonian structures? If the answer is affirmative then is the
mapping $A,B\mapsto AB^{-1}$ Hamiltonian? The fate of the ``first" structure is
not clear, either. We failed to restrict it to $\cR$.

Also notice that the hierarchy (3) has many features in common with the
``modified KP" suggested by Kupershmidt [5] although they are different
(Kupershmidt's modified KP is rather extension than restriction of KP).
Perhaps, there is a more general construction which unites them.\\

{\bf References.}\\

\noindent {\bf 1.} Aratyn, H., Nissimov, E., and Pacheva, S., Construction
of KP hierarchies in terms of finite number of fields and their abelianization,
preprint, hep-th 9306035, 1993\\

\noindent {\bf 2.} Bonora, L. and Xiong, C. S., The (N,M)-th KdV hierarchy
and the associated W algebra, preprint SISSA, hep-th 9311070, 1993\\
Bonora, L., Lin, Q. P., and Xiong, C. S., The integrable hierarchy constructed
from a pair of higher KdV hierarchies and its associated W algebra, preprint,
hep-th 9408035, 1994\\

\noindent {\bf 3.} Dickey, L. A., On the constrained KP hierarchy, preprint,
hep-th 9407038, 1994\\

\noindent {\bf 4.} Dickey, L. A., {\em Integrable equations and Hamiltonian
systems}, Advanced Series in Mathematical Physics, Vol.12, World Scientific,
Singapore, (1991), p.310\\

\noindent {\bf 5.} Kupershmidt, B. A., {\em J. Phys. A: Math. Gen.} 22, L993
(1989)

\end{document}